\input harvmac
\sequentialequations

\def\winf{\omega _{\infty}}
\def\abs{{\cal A}}
\def\scat{{\cal S}}
\def\phinf{\phi _{\infty}}
\def\phmid{\phi _{mid}}

\def\flux{ {\cal F}}

\def\onebar{{\bar 1}}
\def\fivebar{{\bar 5}}


\def\half{{1\over 2}}

\def\sa{{\sinh^2\alpha}}
\def\sg{{\sinh^2\gamma}}
\def\({\left(}
\def\){\right)}
\def\]{\right]}
\def\[{\left[}
\def\winf{\omega_{\infty}}

\def\p{\partial}

\lref\horowitz{G.T. Horowitz, {\it The Origin of Black Hole Entropy in 
String Theory}, to appear in the proceedings of the Pacific Conference 
on Gravity
and Cosmology, gr-qc/9604051 .}
\lref\maldacena{J. Maldacena, {\it Black Holes in String Theory}, 
hep-th/9607235 .}
\lref\polchinski{J. Polchinski, hep-th/9510017 .}
\lref\strominger{A. Strominger and C. Vafa, hep-th/9601029 .}
\lref\cy{M. Cvetic and D. Youm, 
{\it General Rotating, Five Dimensional Black Holes of Toroidally 
Compactified Heterotic String}, 
Nucl. Phys. {\bf B476} (1996) 118, hep-th/9603100 .} 
\lref\dmthree{S.R. Das and S.D. Mathur, {\it Interactions Involving D-Branes},
hep-th/???}
\lref\igor{S.S. Gubser and I.R. Klebanov, {\it Emission of Charged Particles from Four-
and Five-Dimensional Black Holes}, hep-th/9608108 .}
\lref\dkt{H.F. Dowker, D. Kastor and J.Traschen, unpublished.}
\lref\callan{C.G. Callan and J. Maldacena, {\it D-Brane Approach
to Black Hole Quantum Mechanics}, Nucl. Phys. {\bf B472} (1996) 591, hep-th/9602043 .} 
\lref\hs{G. Horowitz and A. Strominger, {\it Counting States of
Near Extremal Black Holes}, hep-th/9602051 .}  
\lref\hms{G.T. Horowitz, J. Maldacena and A. Strominger, 
{\it Nonextremal Black Hole Microstates and U-Duality}, Phys. Lett. {\bf 383B} (1996) 151, 
hep-th/9603109.}
\lref\dmone{S.R. Das and S.D. Mathur, {\it Excitations of D-Strings, Entropy
and Duality}, Phys. Lett. {\bf 375B} (1996) 103, hep-th/9601152 .}
\lref\wadia{A. Dhar, G. Mandal and S.R. Wadia,  {\it Absorption vs. Decay of
Black Holes in String Theory and T-Symmetry}, Phys. Lett. {\bf 388B} (1996) 51, hep-th/9605234 .}
\lref\msuss{J. Maldacena and L. Susskind, {\it D-Branes and Fat Black
Holes}, Nucl. Phys. {\bf B475} (1996) 670, hep-th/9604042 .}  
\lref\dmtwo{S.R. Das and S.D. Mathur. {\it Comparing
Decay Rates for Black Holes and D-Branes}, Nucl. Phys. {\bf B478} (1996) 561, hep-th/9606185 .}
\lref\gibbons{G.W. Gibbons, {\it Vacuum Polarization and the Spontaneous Loss
of Charge by Black Holes}, Comm. Math. Phys. {\bf 44}, 245 (1975).}
\lref\hashimoto{A. Hashimoto and I.R. Klebanov, {\it Decay of Excited D-branes}, 
Phys. Lett. {\bf 381B} (1996) 437, hep-th/9604065 .}
\lref\gubser{S.S. Gubser and I.R. Klebanov, {\it Four-Dimensional Greybody
Factors and the Effective String}, Phys. Rev. Lett. {\bf 77}, 4491, hep-th/9609076 .}
\lref\fixed{C.G. Callan, S.S. Gubser, I.R. Klebanov and A.A. Tseytlin, 
{\it Absorption of Fixed Scalars and the D-brane Approach to Black Holes},
hep-th/9610172 .}
\lref\dealwis{S.P. de Alwis and K. Sato, {\it Radiation from a Class of String
Theoretic Black Holes}, hep-th/9611189 .}
\lref\fixedtwo{I.R. Klebanov and M. Krasnitz, {\it Fixed Scalar Greybody
Factors in Five and Four Dimensions}, hep-th/9612051 .}
\lref\esko{E. Keski-Vakkuri and P. Krauss, {\it Microcanonical D-branes and
Back Reaction}, hep-th/9610045 .}
\lref\hashimotob{A. Hashimoto, {\it Perturbative Dynamics of Fractional Strings
on Multiply Wound D-Strings}, hep-th/9610250 .}
\lref\polhemus{G. Polhemus, {\it Statistical Mechanics of Multiply Wound
D-Branes}, hep-th/9612130 .}

\lref\page{D.N. Page, {\it Particle Emission Rates from a Black Hole: Massless Particles from an 
Uncharged Hole}, Phys. Rev. {\bf D13} (1976) 198.}
\lref\unruh{W.G. Unruh, {\it Absorption Cross Section of Small Black Holes}, Phys. Rev. {\bf D14} 
(1976) 325.}
\lref\star{A.A. Starobinski and S.M. Churilov, {\it Amplification of Electromagnetic and Gravitational 
Waves by A Rotating ``Black Hole''}, Sov. Phys. JETP {\bf 38} (1974) 1.}
\lref\press{S.A. Teukolsky and W.H. Press, {\it Perturbations of a Rotating Black Hole III: 
Interaction of the Hole with Gravitational and Electromagnetic Radiation}, Ap. J. {\bf 193} (1974) 443.}
\lref\hawk{S.W. Hawking, {\it Particle Creation by Black Holes}, 
Comm. Math. Phys. {\bf 43} (1975) 199.}
\lref\ms{J. Maldacena and A. Strominger, {\it Black Hole Greybody Factors and
D-Brane Spectroscopy}, Phys. Rev. {\bf D55} (1997) 861, hep-th/9609026 .} 
\lref\sa{M. Abramowitz and I.A. Stegun, {\it Handbook of Mathematical
Functions}, National Bureau of Standards New York: Wiley (1964).}
\lref\fluids{M. Van Dyke, {\it Perturbation Methods in Fluid Mechanics},
The Parabolic Press, Stanford California (1975).}
\lref\km{I.R. Klebanov and S.D. Mathur, {\it Black Hole Greybody Factors and Absorption of 
Scalars by Effective Strings}, hep-th/9701187.}
\lref\mstwo{J. Maldacena and A. Strominger, {\it Universal Low-Energy Dynamics for Rotating 
Black Holes}, hep-th/9702015.}
\lref\taylor{S.W. Hawking and M.M. Taylor-Robinson, 
{\it Evolution of Near Extremal Black Holes}, hep-th/9702045.}
\lref\dasgupta{S. Das, A. Dasgupta and T. Sarkar, {\it High Energy Effects on D-Brane and 
Black Hole Emission Rates}, hep-th/9702075}
\lref\dgm{S.R. Das, G. Gibbons and S.D. Mathur, {\it Universality of Low-Energy Absorption Cross-Sections
for Black Holes}, Phys. Rev. Lett. {\bf 78} (1997) 417, hep-th/9609052 .}
\lref\larsen{F. Larsen, {\it A String Model of Black Hole Microstates}, hep-th/9702153 .}

\Title{\vbox{\baselineskip12pt
\hbox{UMHEP-437}
\hbox{Imperial/TP/96-97/24}
\hbox{hep-th/9702109}}}
{\vbox{\centerline{\titlerm U-Duality, D-Branes and Black Hole Emission Rates:}
\smallskip
\centerline{Agreements and Disagreements}
 }}
{\baselineskip=12pt
\centerline{Fay Dowker$^{a}$, David Kastor$^{b,1}$
and Jennie Traschen$^{b,2}$  }
\bigskip
\centerline{\sl ${}^a$ Blackett Laboratory,}
\centerline{\sl Imperial College of Science, Technology and Medicine,}
\centerline{\sl London, SW7 2BZ.} 
\centerline{\it Internet: dowker@ic.ac.uk}
\medskip
\centerline{\sl ${}^{b}$ Department of Physics and Astronomy}
\centerline{\sl University of Massachusetts}
\centerline{\sl Amherst, MA 01003-4525}
\centerline{\it ${}^{1}$ Internet: kastor@phast.umass.edu}
\centerline{\it ${}^{2}$ Internet: lboo@phast.umass.edu}}
\medskip
\centerline{\bf Abstract}
An expression for the spacetime absorption coefficient of a scalar
field in a five dimensional, near extremal black hole background is derived,
which has the same form as that presented by Maldacena and Strominger, but is
valid over a larger, U-duality invariant region of parameter space and in general
disagrees with the corresponding D-brane result.  We develop an
argument, based on D-brane thermodynamics, which specifies the range of
parameters over which agreement should be expected.  For neutral emission, the
spacetime and D-brane results agree over this range.  However, for charged
emission, we find disagreement in the `Fat Black Hole' regime,
in which charge is quantized in smaller units on the brane, than in the bulk of
spacetime.  We indicate a possible problem with the D-brane model in this
regime.  We also use the Born approximation to study the high frequency
limit of the absorption coefficient and find that it approaches unity, for large
black hole backgrounds, at frequencies still below the string scale, again in
disagreement with D-brane results.

\Date{February, 1997}
\vfill
\eject

\newsec{Introduction}  
Progress made in the past year has led towards a possible
understanding of the nature of black hole microstates within string theory
(see {\it e.g.} the reviews \refs{\horowitz,\maldacena} and references therein).  Counting
arguments based on the weak coupling D-brane description of string solitons 
have been shown to
reproduce the Bekenstein-Hawking formula for the entropy of certain extremal and
near-extremal black holes.   The D-brane description also
leads to a manifestly unitary prescription for calculating the rate of 
absorption/emission from the D-brane system \callan ,
reproducing the characteristic black body form of the Hawking emission spectrum.
A number of authors
\refs{\wadia,\dmtwo,\hashimoto,\ms,\gubser,\fixed,\dealwis,\fixedtwo,\esko,
\km,\mstwo,\taylor,\dasgupta} 
have gone on to
compare in more detail the spectra of emitted radiation calculated for the
D-brane system with that calculated using semiclassical spacetime methods.

In the present paper we continue this work of detailed comparison between the
D-brane and spacetime calculations.  We present a number of new results, 
including a U-duality invariant (covariant) expression for the spacetime emission rate of neutral
(charged) particles, together with a discussion based on D-brane thermodynamics of
the regime in which we would expect to obtain agreement 
between this spacetime expression and D-brane results; a discussion of
the expected higher energy behavior of the spacetime emission rate; and, most interestingly,
a regime of disagreement between the D-brane and
spacetime calculations for emission of charged particles. This regime of disagreement occurs in the
`fat black hole' region of parameter space, in which charge is quantized in smaller units on the
brane, than in the bulk of spacetime \refs{\dmone,\msuss}.  
It is left as an open question whether this disagreement
can be resolved through a clearer understanding 
of emission processes in the `fat black hole' limit.

One might be tempted to dismiss these discrepancies as irrelevant since 
it was argued in \refs{\ms} that the spacetime calculation of emission 
from a black hole is only valid for black holes larger than the string scale
whereas the perturbative D-brane calculation is only valid for black holes
smaller than the string scale. This would mean that there should be no reason 
to expect agreement between the two. Nevertheless, agreement, where it has 
been found, has been taken as further evidence for the general idea that 
D-branes provide the correct microscopic description of black holes in 
string theory. It is important, then, not only to point out when unexpected
agreement occurs, but also to identify when the calculations of emission 
rates, say, actually differ. This information  
may be just as valuable in elucidating the black hole/D-brane 
correspondence.

The spectrum of energy emitted in Hawking radiation \hawk\ by a charged black hole is
given by 
\eqn\blackbody{
{dE\over dt}(\omega, q)={\omega|\abs (\omega,q)|^2\over 2\pi
\left(e^{ (\omega -q\Phi _e)/T_H}-1\right)}, }
where $\omega$ and $q$ are the energy and charge of the emitted particle, 
$T_H$ is the Hawking temperature and the
chemical potential $\Phi_e$ is the difference in the electrostatic potential 
between infinity and the black hole horizon, 
$\Phi_e = \Phi_{\infty} -\Phi _h$.  The prefactor $|{\cal A}(\omega, q)|^2$,
required by detailed balance, is the classical absorption coefficient for the
emitted mode.  

In the case of neutral emission, $q=0$ in \blackbody, 
Das and Mathur \dmtwo\ showed agreement between the D-brane and
spacetime calculations of the absorption
coefficient $\abs (\omega)\equiv {\cal A}(\omega,q=0)$ at leading order in a small $\omega$
approximation. Maldacena and Strominger \ms\ then derived a more detailed result
for the spacetime calculation of $\abs (\omega)$, showing striking detailed
agreement with the D-brane result of \dmtwo (though still in a low energy
approximation).  Below we show that a result for the spacetime
absorption coefficient $\abs (\omega)$ of the form given in \ms\ continues to hold over 
a larger region
of black hole parameter space, yielding an expression which is invariant
under U-duality transformations. The expression is symmetric under interchange of
the three charges carried by  the black hole, as one should expect for neutral
emission.  

This new expression for $\abs (\omega)$ disagrees in general with the
D-brane result.  We then ask the question, given that calculations in the
D-brane system are limited to processes in which the dynamics of branes and anti-branes
are unimportant, over what region of parameter space should we expect the
D-brane result to be accurate?  We show that if one assumes that the statistical
D-brane system is described by the U-duality invariant entropy formula given in
\hms , as it must be to correspond to the black hole, then the region of
parameter space, for which left and right moving excitations dominate the process
of energy exchange of the system with its environment, matches precisely 
the regime for which the D-brane and spacetime results for $\abs (\omega)$
agree.  This demonstrates a degree of consistency in the assumption that a full
understanding of D-brane dynamics would reproduce the U-duality invariant
results.  
The full expressions, moreover, may give clues towards such an understanding.
For example, in the U-duality related limit in which 1-branes and anti-1-branes dominate, the
expression for $\abs (\omega)$ may be interpreted in terms of distributions of branes and anti-branes.


Turning to the case of charged emission, $q\ne 0$ in \blackbody, contrary to
the conclusions of \refs{\igor,\ms} , we find real disagreement between the spacetime
and D-brane results for ${\cal A}(\omega, q)$ in an interesting region of
parameter space.   For charged emission, the spacetime
result was first calculated and compared with the D-brane result \refs{\igor,\dkt} to leading
order in the small parameter 
\eqn\smallparam{\omega^2 -q^2.}   
A more detailed result for the spacetime absorption coefficient was given in
\ms , again showing apparent striking agreement with the D-brane result. We show here that this
agreement does not extend over the entire parameter range of interest.  
In particular, it does not
hold in the `fat black hole' (hereafter FBH) regime \refs{\dmone,\msuss}.
We find that there are two cases to consider; Case I, in which the difference \smallparam\ is
small, with both $\omega$ and $q$ individually small; and Case II, in which
the  difference \smallparam\ is small, but neither $\omega$ nor $q$ is small.
We find that the D-brane and spacetime results for ${\cal A}(\omega,q)$ agree
only in case I.  Case II, however, applies in an interesting region of parameter
space, the FBH regime.  It is left as an open
question, whether the lack of agreement we find between the spacetime and
D-brane results in this case shows a real disagreement or an error in the
D-brane result.  Such an error could arise in the FBH regime because charge is
stored on the brane in smaller units than may be carried by excitations in the
bulk of spacetime \dmone\msuss.  Only very special combinations of brane
excitations can actually annihilate to carry charge off the brane via the simple
interactions considered in \dmtwo .


Thirdly, it should be recognized that the emission rate which has
been computed for the D-brane system does not by any means hold
over the entire black body spectrum, for either neutral or charged
emission.  In particular, as we will discuss below,
in the high frequency limit the black hole absorption coefficient approaches
unity. This is just the particle limit - an incident, zero angular 
momentum particle is absorbed by the black hole with unit probability. 
For large black holes,
the characteristic frequency $\omega _{over}$
at which the absorption coefficient goes
to one, though well above the Hawking temperature,
is still small compared to the string scale. So at least naively,
one should be able to use the effective vertex for
weakly  interacting left and right movers given in \refs{\hashimoto,\dmtwo} 
in this regime.


This paper is organized as follows:  In section 2 we review the properties of
the family of spacetimes and the D-brane model, including the D-brane result for the absorption 
coefficient. In section 3, we establish, via thermodynamic arguments based on the self-consistency of 
the D-brane model, the regime in which we should expect agreement between the D-brane and 
spacetime results.  Section 4 presents the calculation of the spacetime absorption coefficient.
Section 5 is a comparison between the spacetime and D-brane results.  Section 6 contains a discussion of the high energy limit of the absorption coefficient.  We present our conclusions in section 7.

\newsec{Black Holes and D-branes}
\subsec{The Spacetimes}
The $5$-dimensional black hole solutions of interest \refs{\hms,\cy} are obtained from 
$10$-dimensional, boosted, black brane solutions to the low energy effective
action of Type IIB string theory, which contains the terms (in the $10$d
Einstein frame)
\eqn\action{
{1\over 16\pi G_{10}}\int d^{10} x \sqrt{-g} \left( R - \half (\nabla\phi)^2
-{1\over 12} e^{\phi} H^2 \right), }
where $H$ is the RR $3$-form, $\phi$ is the dilaton, $G_{10} = 8\pi^6 g^2$,
$\alpha' =1$ and $g$ is the $d=10$ string coupling. The $d=10$ solutions are
given by 
\eqn\tensol{ 
\eqalign{
ds^2 = &
 \( 1 + { r_0^2 \sinh^2\alpha\over r^2}\)^{-3/4} \( 1 + {r_0^2 \sg \over r^2}\)^{-1/4}
\left[ - dt^2 +dx_5^2
\right. \cr
+& \left. {
r^2_0  \over r^2} (\cosh \sigma dt - \sinh\sigma dx_5)^2
 +\( 1 + {r_0^2\sinh^2\alpha \over r^2}\) \sum_{i=6}^{9}dx_i^2 \] \cr
 +& \( 1 + { r_0^2\sinh^2\alpha \over r^2}\)^{1/4}
\( 1 + { r_0^2\sg \over r^2}\)^{3/4} \left[
\(1-{r_0^2 \over r^2}\)^{-1} dr^2 + r^2 d \Omega_3^2 \right]\cr
e^{-2\phi} = &\(1+ {r_0^2\sg\over r^2}\)\(1 + {r_0^2\sinh^2\alpha\over r^2}\)^{-1}\cr
H=&2r_0^2\sinh^2\gamma\epsilon_3 +2r_0^2\sinh^2\alpha
e^{-\phi}*_6\epsilon_3,\cr }}
where $*_6$ is the Hodge dual in the six dimensions $t,\dots,x_5$
and $\epsilon_3$ is the volume element on the $3$-sphere.
The $x_5$ coordinate is taken to be compact with period $2\pi R$.  The
coordinates $x_6,\dots,x_9$ are also taken to be compact and are each identified
with period $ 2\pi V^{1/4}$. The $d=10$ solution is then specified by the six
parameters $\alpha,\gamma,\sigma, r_0, R,V$.  These may be exchanged for a set
of six physical parameters; three charges $Q_1,Q_5,n$, the ADM energy $E$, $R$
and $V$.  The three charges are given by
\eqn\charges{
\eqalign{
   Q_1 &= {V\over 4\pi^2 g}\int e^\phi *_6H
   = { V r_0^2  \over 2 g } \sinh 2 \alpha , \cr
   Q_5 &= {1\over 4\pi^2 g} \int H  =  { r_0^2\over 2g} \sinh 2 \gamma ,
\cr
  n &= {  R^2V r_0^2 \over 2  g^2} \sinh 2 \sigma\ . \cr
}}
The third charge $n$ is related to the  momentum around the circle in the $x_5$
direction by $P=  n/R$. 
The ADM energy is 
\eqn\energy{E={RVr_0^2\over 2g^2}(\cosh 2\alpha+\cosh 2\gamma +
\cosh 2\sigma).} 
The black hole entropy $S$, Hawking temperature $T_H$ and chemical potential $\mu_n$ (related by a factor
$1/R$ to $\Phi_e$ above) 
are given by
\eqn\hawking{S={2\pi RVr_0^3\over
g^2}\cosh\alpha\cosh\gamma\cosh\sigma,\qquad {1\over T_H}=2\pi
r_0\cosh\alpha\cosh\gamma\cosh\sigma, \qquad \mu_n={\tanh\sigma\over R}.}

The reduced five-dimensional metric and additional Kaluza-Klein moduli
can be read off from 
\eqn\reduced{
ds_{10}^2= e^{2\chi}\sum_i dx_i^2 + e^{2\psi}
(dx_5+A_\mu dx^\mu)^2 + e^{-2(4\chi + \psi)/3} ds_5^2}
where $\mu=0,1,...4$. Using these normalizations, $ds_5^2$ is the 
Einstein metric in five dimensions, which for \tensol\ takes the simple form
\eqn\fivesol{\eqalign{ds_5^2&=-f^{-2/3}\left(1-{r_0^2\over r^2}\right)dt^2
+f^{1/3}\left[\left(1-{r_0^2\over r^2}\right)^{-1}dr^2+
r^2d\Omega_3^2\right],\cr 
f&=\left(1+{r_1^2\over r^2}\right)\left(1+{r_5^2\over r^2}\right)
\left(1+{r_n^2\over r^2}\right),\cr}}
where the characteristic radii 
$r_1=r_0\sinh\alpha$, $r_5=r_0\sinh\gamma$ and $r_n=r_0\sinh\sigma$ are 
sometimes referred to as `charges' below.

\subsec{The D-brane Model}
In reference \hms\ the six physical parameters above were further exchanged for
a set of six `brane numbers' via the relations
\eqn\branenumbers{\eqalign{
Q_1& =N_1-N_\onebar,\qquad Q_5=N_5-N_\fivebar,\qquad n=N_L-N_R,\cr
E & ={R\over g}(N_1+N_\onebar) +{RV\over g}(N_5+N_\fivebar)+{1\over R} 
(N_R+N_L), \cr
V& =\left({N_1N_\onebar\over N_5N_\fivebar}\right)^{1/2},\qquad 
R=\left({g^2N_RN_L\over N_1N_\onebar}\right)^{1/4}.\cr }}
$N_{1}$ and $N_{\bar{1}}$ are thought of as the
numbers of 1D-branes and  anti-1D-branes wrapping around the circle in the 
$x_5$-direction.  Likewise, 
$N_5$ and $N_{\bar{5}}$ are thought of as the numbers of 5D-branes and
anti-5D-branes wrapping around the internal $5$-torus, and 
$N_L/R$ and  $N_R/R$ are the total momenta carried by massless, left and right
moving, open string excitations propagating along the common $x_5$
coordinate of the branes.  In terms of the boost parameters, $\alpha,\gamma,\sigma$
appearing in the metric \tensol, the brane numbers
are given by
\eqn\boosts{\eqalign{ N_1&={Vr_0^2\over 4g}e^{2\alpha},\qquad
N_\onebar={Vr_0^2\over 4g}e^{-2\alpha},\cr
N_5&={r_0^2\over 4g}e^{2\gamma},\qquad 
N_\fivebar={r_0^2\over 4g}e^{-2\gamma},\cr
N_L&={r_0^2R^2V\over 4g^2}e^{2\sigma},\qquad 
N_R= {r_0^2R^2V\over 4g^2}e^{-2\sigma}.\cr}}
The black hole entropy, expressed in terms 
of the brane numbers, takes the simple form \hms\
\eqn\entropy{S= 2\pi \left(\sqrt{N_1}+\sqrt{N_{\bar 1}}\right)
\left(\sqrt{N_5}+\sqrt{N_{\bar 5}}\right)
\left(\sqrt{N_R}+\sqrt{N_L}\right).    }

At this level, the brane numbers represent only a relabeling of parameters. 
However, in the near extremal, weak string coupling limit, a string theory
based counting argument, in which the brane numbers stand for actual numbers of
branes, reproduces the appropriate limit of \entropy\ as the statistical
degeneracy of the system \refs{\callan,\hs,\hms}.  Very briefly, the construction is the following.
If the charges $Q_1,Q_5,n$ are all taken to be
positive, then the extremal limit is $N_\onebar=N_\fivebar=N_R=0$.  Moving
slightly away from extremality, the expression for the energy $E$ in \branenumbers\
indicates that, if $R^2/g,R^2V/g\gg 1$, momentum modes will be light
compared to anti-branes.  Adding a small amount of energy
will then cause $N_R$  to increase, with $N_\onebar,N_\fivebar$
remaining approximately zero.  The counting arguments then proceed by counting
the number of distinct configurations of left and right moving open string
excitations having the correct total momentum, $N_L$ and $N_R$.

In order correctly to reproduce the limiting form of the entropy \entropy\ in the 
extremal limit, 
$N_\onebar=N_\fivebar=N_L=0$, it is necessary to assume that the momenta of
open string excitations propagating along the brane are quantized in units of
$1/L$, with $L=N_1N_5R$, rather than in units of $1/R$ as for closed string
states propagating in the bulk of the spacetime \refs{\dmone,\msuss}.  
We refer to this as the `fat
black hole' (hereafter FBH) prescription.  How the FBH prescription arises from
string theory has been explored recently in \refs{\hashimotob,\polhemus}.

Moving away from extremality by adding right-movers, keeping $N_\onebar,N_\fivebar\simeq 0$, 
in the limit $N_L,N_R\gg 1$, the microcanonical distributions of massless, left
and right moving, open string excitations may be replaced by canonical
distributions,
\eqn\canonical{
\rho_L(\omega_L) ={1\over e^{\omega_L/ T_L }-1},\qquad
\rho_R(\omega_R) ={1\over e^{\omega_R/ T_R }-1}, \qquad}
The temperatures $T_L,T_R$, determined by the conditions that the average total
momenta carried by the left and right movers be $N_L/R,N_R/R$, are given by
\eqn\temperatures{\eqalign{{1\over T_L}&=\pi R\sqrt{{N_1N_5\over N_L}}
={\pi r_1r_5\over 2r_0^2}\left(\sqrt{r_n^2+r_0^2}-r_n\right),\cr
{1\over T_R}&=\pi R\sqrt{{N_1N_5\over N_R}}
={\pi r_1r_5\over 2r_0^2}\left(\sqrt{r_n^2+r_0^2}+r_n\right).\cr}}
$T_R$ vanishes in the extremal limit.  The entropy \entropy\ with
$N_\onebar=N_\fivebar =0$ is then reproduced by the sum of two non-interacting
one dimensional ideal gasses,
\eqn\onedgas{S=S_L+S_R=2\pi^2(T_L+T_R)L.}

The excited D-brane system, {\it i.e.} with $N_R>0$, 
decays via closed string emission as described in \callan .
A left moving open string having energy $\omega_L$ annihilates with a right moving open
string having energy $\omega_R$ to form a closed string having energy 
$k_0=\omega_L+\omega_R$ and
internal momentum $k_5=\omega_L-\omega_R$.  From a $5$-dimensional 
point of view, the
closed string state carries electric charge $q=k_5$.  Recall that the internal
momentum $k_5$ of a closed string state is quantized in units of $1/R$, whereas
the $\omega_L,\omega_R$ are quantized in units of $1/L=1/N_1N_5R$.  This implies a strong
restriction on which left and right movers may annihilate to form a closed
string which can propagate into the bulk spacetime via this interaction.

The basic interaction vertex for the above process has been determined both
from the low energy limit of perturbative string calculations \hashimoto\ and
from the Born-Infeld effective action for the D-brane system \dmtwo . The
rate for both neutral and charged emission can be calculated using the methods
of \dmtwo\ giving the result \refs{\igor,\dkt}
\eqn\rate{
{dE\over dt}(\omega,k_5)  =  \omega G_5{N_1 N_5 R\over 2 }
 (\omega^2 -k_5^2)^2 \rho_L(\omega_L )
\rho_R(\omega_R )      ,}
where $G_5=\pi g^2/4RV$.
Comparing this expression with \blackbody , the D-brane prediction for the absorption
coefficient is then
\eqn\dbraneabsorp{|{\cal A}_D(\omega,k_5)|^2=\pi G_5 N_1N_5R(\omega^2-k_5^2)
{e^{(\omega-R\mu_n k_5)/T_H}-1\over (e^{(\omega+k_5)/2T_L}-1)(e^{(\omega-k_5)/2T_R}-1)}}

\newsec{Expected Regime of Agreement Between D-brane and Spacetime Results}
\subsec{Neutral Emission}
In section (4) we will obtain a result for the spacetime absorption
coefficient $|{\cal A}(\omega,k_5)|^2$ which disagrees in general with the D-brane result 
$|{\cal A}_D(\omega,k_5)|^2$ in \dbraneabsorp\ above.  We would like to ask, in advance,
given that the D-brane calculation does not take into account brane/anti-brane
dynamics, over what regime of black hole parameter space should we expect the
D-brane and spacetime results to agree?

For neutral emission ($k_5=0$), the D-brane
system is exchanging a small amount of energy with its environment, the
other physical parameters being held fixed.  If the D-brane system does indeed correspond
to the black hole, then we can use \branenumbers\ to calculate how the 
brane numbers change under a small change in the energy $E$, with the
other physical parameters $Q_1,Q_5,n,R$ and $V$ held fixed.  We find these
partial derivatives of the brane numbers to be given by
\eqn\variations{\eqalign{ {\partial N_1\over \partial
E}&= {\partial N_\onebar\over \partial E}={g\cosh 2\gamma\cosh 2\sigma\over
2R\Omega}\cr
{\partial N_5\over \partial E}&=
{\partial N_\fivebar\over \partial E}= 
{g\cosh 2\alpha\cosh 2\sigma\over 2RV\Omega}\cr
{\partial N_R\over \partial E}&=
{\partial N_L\over \partial E}= {R\cosh 2\alpha\cosh 2\gamma\over 2\Omega},\cr}}
where 
\eqn\factor{\Omega=\cosh 2\alpha\cosh 2\gamma +\cosh 2\alpha\cosh 2\sigma
+\cosh 2\gamma\cosh 2\sigma .}
The dependence on $g,R,V$ in these expressions reflects the
masses of the different brane species, as discussed above.  The dependence on the
boost parameters $\alpha,\gamma,\sigma$, arises from holding $R,V$ fixed.  We see
that for the case of equal charges, $\alpha=\gamma=\sigma$, the change in a
given brane number is simply the inverse of its mass per unit excitation.

The partial derivatives \variations\ are ingredients in a calculation of the
Hawking temperature in terms of the brane numbers.  We can write
\eqn\temperature{{1\over T_H}= 
{\partial S\over \partial E}
=\sum_i\left( {\partial N_i\over \partial E}\right){\partial S\over \partial N_i}
} 
where the index $i$ runs over the six types of excitations.  The derivatives 
$\partial S/\partial N_i$ are easily calculated from \entropy\ and we arrive at
\eqn\branetemp{
{1\over T_H}={2\pi r_0\cosh\alpha\cosh\gamma\cosh\sigma\over\Omega}
\left\{ \cosh 2\gamma\cosh 2\sigma +
\cosh 2\alpha\cosh 2\sigma  +
\cosh 2\alpha\cosh 2\gamma\right\},}
which correctly reproduces the Hawking temperature \hawking\ of the black hole,
since the three terms in brackets sum to $\Omega$.  The three individual terms in
\branetemp\ come respectively from 1-branes \& anti-1-branes,
5-branes \& anti-5-branes and left \& right moving excitations.  We see that the relative
contributions have the same dependence on the boost parameters as appear in
\variations .  However, the factors of $g,R,V$ arising from the masses of the
different excitations have been normalized away by the coefficients  $\partial
S/\partial N_i$.

We suggest that it is the relative size of the contribution of each
brane species to the Hawking temperature, as in \branetemp , which 
determines its importance to processes of energy exchange, rather than the change
in brane number itself.  For example, with equal charges, 
$\alpha=\gamma=\sigma$,
all three sets of excitations contribute equally to the temperature and are therefore
equally important thermodynamically.
By this criterion, we see from \branetemp, using the expression for $\Omega$ 
\factor, that left and right moving
excitations dominate the expression for $T_H$ in the limit $\sigma\ll\alpha,\gamma$, which
is equivalent to the condition 
\eqn\mslimit{r_n\ll r_1,r_5.}  
If this criterion is correct, it is in this regime that we should expect
agreement between the D-brane and spacetime results.  Note that the region
of parameter space satisfying \mslimit\ is the same as the `dilute gas' limit 
in reference \ms .  In \ms , equation \mslimit\ arose as the consistency
condition for a small amplitude approximation of the dynamics of classical waves on a
string.  Here we see that the same condition emerges from considerations specific to the
D-brane system under study.  Moreover, in this way of deriving the condition
\mslimit , it is explicit that in the limit \mslimit\ the inverse Hawking temperature reduces 
as in \ms\ to
\eqn\TH{{1\over T_H}={1\over 2}\left( {1\over T_L}+{1\over T_R}\right).}

It is also clear from the above analysis that in the regimes related by 
U-duality to \mslimit , {\it i.e.} $r_1\ll r_n,r_5$ and $r_5\ll r_1,r_n$, the
dynamics will be dominated by 1-branes and 5-branes respectively.  In the limit 
\eqn\onelimit{r_1\ll r_n,r_5,} 
for example, the Hawking temperature is approximately
\eqn\TH{{1\over T_H}={1\over 2}\left( {1\over T_1}+{1\over T_\onebar}\right),}
where 
\eqn\onetemp{ {1\over T_1}={2\pi r_5r_n\over r_0^2}\left(
\sqrt{r_0^2+r_1^2}-r_1\right),\qquad
{1\over T_\onebar}={2\pi r_5r_n\over r_0^2}\left(
\sqrt{r_0^2+r_1^2}+r_1\right)}
can be thought of as the temperatures of canonical distributions of 1-branes
and anti-1-branes.  These same temperatures will arise in our expression for
the spacetime absorption below after taking the limit \onelimit .

\subsec{Charged Emission}
Our result for the spacetime absorption
coefficient $|{\cal A}(\omega,k_5)|^2$ with $k_5\neq 0$
disagrees as well in general with the 
D-brane result \dbraneabsorp .  So, again we would like to ask over what region
of parameter space we should expect agreement.  We can perform an analysis 
similar to
that above, but now for a process in which the energy and charge change in a
fixed ratio. To understand which excitations dominate the emission of charge, we
calculate the chemical potential $\mu_n$, for emitting particles carrying
$n$-charge in terms of the brane numbers,
\eqn\chemical{
{\mu_n\over T}=  - {\partial S\over \partial n}
=  -\sum_i \left({\partial N_i\over\partial n}\right)
\left({\partial S\over\partial N_i}\right) ,}
where the partial derivatives are taken holding $E,Q_1,Q_5,R$ and $V$ constant.
We find the derivatives of the brane numbers are
\eqn\chargedvar{\eqalign{
{\partial N_R\over\partial n}&={1\over 2\Omega}\left(\cosh 2\alpha\cosh 2\gamma
+\cosh 2\alpha e^{2\sigma} +\cosh 2\gamma e^{2\sigma} \right),\cr
{\partial N_L\over\partial n}&={1\over 2\Omega}\left(\cosh 2\alpha\cosh 2\gamma
+\cosh 2\alpha e^{-2\sigma} +\cosh 2\gamma e^{-2\sigma} \right),\cr
{\partial N_1\over\partial n}&={\partial N_\onebar\over\partial n}=
-{g\cosh 2\gamma\sinh 2\sigma\over 2R^2\Omega},\cr
{\partial N_5\over\partial n}&={\partial N_\fivebar\over\partial n}=
-{g\cosh 2\alpha\sinh 2\sigma\over 2R^2\Omega},\cr}}
Assembling these into a calculation of the chemical potential $\mu_n$ gives
\eqn\chempot{\eqalign{
{\mu_n\over T_H}={2\pi r_0\cosh\alpha\cosh\gamma\cosh\sigma\over\Omega}
&\left\{2\cosh 2\gamma\cosh^2\sigma +2\cosh 2\alpha\cosh^2\sigma\right.\cr
&\left .-\cosh 2\gamma -2\cosh 2\alpha +\cosh 2\alpha\cosh
2\gamma\right\}{\tanh\sigma\over R}.\cr}}
The terms in brackets sum to $\Omega$ giving the correct result
$\mu_n=\tanh\sigma/R$ as in \hawking.  In \chempot\ the first two terms come from
1-branes \& anti-1-branes and 5-branes \& anti-5-branes respectively.  The remaining
terms all come from right \& left moving excitations.  Again, in the limit
$\sigma\ll\alpha,\gamma$, corresponding to the limit \mslimit , we see that the
contributions of right/left moving excitations dominate.  If we now consider a
process in which the energy changes by $k_0$ and the charge $n$ by $Rk_5$, we
can see that the most important processes will be those involving right/left
movers in the regime \mslimit .  So, again we should expect to find agreement
between D-brane and spacetime results in the regime $r_n\ll r_1,r_5$.  

\newsec{Calculating the Spacetime Absorption Coefficient}
\subsec{The Wave Equation}
We now turn to the calculation of the spacetime absorption coefficient.
Consider a massless scalar field $\Phi$ in ten dimensions, minimally coupled to
the ten dimensional metric.  The field is taken to be spherically symmetric,
have frequency $\omega$, momentum $k_5$ in the $x_5$ and to be independent of the
other compact directions.  From a five dimensional point of view, the field
will then carry charge $q=k_5$. Let 
\eqn\scalar{\Phi = e^{-i\omega t} e^{i k_5 x_5} \varphi(r)}
where 
$k = m/R$ with $m$ an integer. 
The ten dimensional Klein-Gordon equation becomes
\eqn\wave{ 
 {\(1-{r_0^2\over r^2}\) \over r^3 }
\p_r\[r^3\(1 - {r_0^2\over r^2}\) \p_r \varphi\] +
\[{\winf}^2 +{r_n ^2\over r^2}\mu^2 \]\(1+{r_1 ^2\over r^2}\)
\(1+{r_5 ^2\over r^2}\)f(r)
\varphi = 0
}
where $f(r)$ is as in \fivesol , $\winf^2= \omega^2 - k_5^2$ is the frequency squared
of the wave at spatial infinity and $\mu = \omega - k_5(1+ r_0^2/r_n ^2)^\half$. 
In the extremal limit $r_0 =0$, $\mu$ is the frequency of the wave near the
horizon.

Transforming to the new radial coordinate 
$v={r_0 ^2 \over r^2}$, the wave equation \wave\  becomes
\eqn\waveinv{(1-v){d\over dv}\left( (1-v) \phi '(v) \right) +
\left[ D +{C\over v} +{C_2 \over v^2 } +{C_3 \over v^3 } \right]\phi =0 }
where the constant coefficients $C,C_1,C_2$ and $D$ are given by
\eqn\decoeff{\eqalign{
&C_2 ={1\over 4}\left( \winf ^2 (r_1 ^2 +r_5 ^2 ) +\mu ^2 r_n ^2 \right),
\qquad C_3 ={1\over 4}r_0 ^2 \winf ^2 \cr
&C  ={1 \over 4 r_0^2}\left(  \omega_\infty ^2 r_1^2r_5^2
 + \mu ^2 (r_1^2r_n^2 +r_5^2r_n^2 ) \right),
\qquad D={\mu^2\over 4r_0^4} r_1^2r_5^2r_n^2 ,\cr} }
For vanishing internal momentum $k_5$, $\mu=\omega_\infty$ and all four
coefficients are symmetric under interchange of $r_1,r_5$ and $r_n$.

The scattering problem we consider below is of the same form as that considered
by Maldacena and Strominger \ms .  However, our treatment will differ from that in \ms\
in two important respects.  First, 
we will not make the restriction $r_n\ll r_1,r_5$.
This turns out to be unnecessary, and, consequently,
we arrive at a result for the spacetime absorption coefficient valid over a
larger, U-duality invariant region of parameter space.  Secondly, we treat the
matching problem involved in calculating the absorption coefficient more
carefully, extracting limits on the range of parameters over which the
matching is a good approximation.

Specifically, we find that the spacetime absorption coefficient for general
choices of $r_1,r_5,r_n$ continues to have the form found in \ms
\eqn\clabs{ |{\cal A}(\omega,k_5) |^2 =\omega_\infty ^2 r_0 ^2 \pi ^2 ab
{e^{2\pi (a+b)} -1 \over (e^{2\pi a} -1 ) (e^{2\pi b} -1 ) }
,}
where
\eqn\random{ \qquad   a=\sqrt{C+D} +\sqrt{D}  ,\qquad  b=\sqrt{C+D} -
\sqrt{D} }
and the coefficients $C$ and $D$ are as in \decoeff .  This result 
is valid with the frequency at infinity bounded by the condition.
\eqn\cond{
\winf r_{max} \leq {r_0 \over r_{max} }\ll 1 ,}
where $r_{max}$ is the largest of $r_1,r_5,r_n$ and $r_0$ is restricted by 
$r_0^3\ll r_1r_5r_n$.  Note that 
$\omega_\infty=\sqrt{(\omega-k_5)(\omega+k_5)}$ may be small either because both
$\omega$ and $k_5$ are small, or because one of the factors $\omega\pm k_5$ is
small.  This will be important later on.

Two different kinds of expansion techniques have
been used recently to calculate the absorption coefficient.
One approach, developed in earlier work by  Unruh \unruh\ and Page \page,
and applied in the present context in \refs{\wadia,\dmtwo,\igor,\dkt}
is to note that the wave equation \wave\ is exactly solvable
for $\winf =\mu =0$ (equivalently, $\omega=k_5=0$). The solution 
$\phi (\winf,\mu ;r )$ with finite parameters can then be expanded in a double
power series in $\winf$ and $\mu$. However, this
expansion is valid only in a ``middle region'', neither too close
to the horizon, nor to infinity.  This is  
because the zeroth order solution is singular on the horizon, and does
not approach a plane wave at infinity. One then properly works with
three regions. 
In the tortoise coordinate, the solutions are asymptotically Bessel
functions near infinity and plane waves near the horizon.  These pieces
can be linked together by matching with the low frequency expansion
in between.  The result for the absorption coefficient is then given in terms of a 
power series expansion for small $\winf,\mu$.
 
A second type of expansion, which yields results beyond a power series expansion
for the absorption coefficient, was developed for rotating and charged four
dimensional black holes in \refs{\star,\press,\gibbons}. In these cases the
wave equation can be solved asymptotically both near the horizon and near
infinity in terms of special functions (hypergeometric and confluent
hypergeometric functions), {\bf and}  there is an overlap region in which the
asymptotic forms of the solutions can be matched.  The existence of this overlap
region requires  some parameter to be small, for example $\omega -m\Omega _H$
in the rotating case, where $\Omega_H$ is the angular velocity of the horizon
and $m$ is the angular momentum of the scattered wave.  
A similar approach the present context was introduced in 
\ms\ to obtain results of the form \clabs. In this case the solutions near
infinity and near the horizon are Bessel functions and hypergeometric
functions, respectively.  However, in this case, as we will see below, there
turns out to be no overlap region in which one can match the two asymptotic
regimes. Therefore one needs again to
introduce a ``middle region'', and match the three parts of the solution,
which we do below.

\subsec{Near Infinity} 
The wave equation \waveinv\ becomes Bessel's equation in the regime where 
the terms with coefficients
$C$ and $D$ may be dropped relative to those with coefficients $C_2 ,C_3$. 
Such an approximation
is valid for $r\gg r_{max}$, where again $r_{max}$ is the largest of
$r_1,r_5$ and $r_n$.
Normalizing the incoming part of the wave to unit amplitude, the solution 
$\phi_\infty$ near infinity is given in terms of the $r$ coordinate by
\eqn\phinfin{\phinf (r ) = \sqrt{{\winf \pi \over 2}}
{e^{-i\pi/4}\over r} \left( H_\nu ^{(2)} (\winf r) +i\scat H_\nu ^{(1)}(\winf r)
\right),\quad \nu=1-\omega_\infty^2(r_1^2+r_5^2+r_0^2)+\mu^2r_n^2. }
Here $\scat$ is the scattering coefficient, related to the absorption
coefficient $|{\cal A}|^2$ by the unitarity relation $|\scat |^2 +|\abs |^2 =1 $.  The goal
of the calculation then is to determine $\scat$ given the boundary condition that at the horizon
the outgoing part of the wave vanishes.
Equation \phinfin\ is a large $r$ (small $v$) solution, however the Bessel functions may
still be expanded for small argument, provided 
\eqn\freqsmall{\winf r_{max} \ll 1}
In terms of $v$, this expansion yields the leading terms
\eqn\phinfsmal{ \phinf \sim b_1 v +b_2 +b_3 \log v ,\qquad v\gg \omega_\infty^2 r_0^2}
where the coefficients $b_i$ depend on $\scat$
and $\winf $ and are given in the appendix.

\subsec{Near the Horizon}
The near horizon regime is defined by keeping the $C$ and $D$ terms 
in the wave equation \waveinv\ and dropping the $C_2$ and $C_3$ terms. Several conditions are required
for this approximation to be valid and different limiting
cases are important. One necessary condition in all cases is 
\eqn\rzero{r_0^3\ll r_1r_5r_n.} 
Over the range $r_1\sim r_5$, $r_n\le r_1$ (which includes both the equal charge case and $r_n\ll r_1$),
 a second condition deduced from \waveinv,\decoeff\ 
is $r\ll \sqrt{r_nr_1}$.
The solution $\phi_h$ in this regime, with the boundary condition that there
is no outgoing flux at the horizon is the 
hypergeometric function \ms 
\eqn\phhor{ \phi _h =A (1-v)^{-i(a+b)/2} F( -ia,-ib,1-ia-ib , 1-v) , }
where the constants $a$ and $b$ have been defined in terms of the parameters of the wave equation in 
\random , and
the constant $A$ is to be determined through the matching procedure.
This has the expansion as $v\rightarrow 1$
\eqn\phhorin{\phi _h \sim A \exp \{-i\sqrt{C+D} \log(1-v) \} \ \ 
,\ v\rightarrow 1. }
Expanding $\phi _h$ for $v\rightarrow 0$, away from the horizon, one
finds the leading terms
\eqn\phhorlarg{ \phi _h \sim AE (1+ gv -ab v\log v).
}
Here the function $E(a,b)$ was defined in \ms\ and is given in (A.4)
the appendix below, and the constant $g$ is given by
\eqn\ggg{g = {i\over 2}(a+b) +ab \left( 1 -2\gamma -\psi (1-ia)
-\psi (1-ib) \right), }
where $\psi$ is the digamma function and $\gamma =-\psi (1)$ is Euler's
constant. For the expansion \phhorlarg\ to be valid, with the terms ordered as written 
(which will prove to be important in the matching below), 
$v$ must satisfy
\eqn\hypcond{ abv |\log v|\ll |g|v\ll 1. }
We find by plugging in the actual coefficients $a,b,g$ that these inequalities require 
$v\gg \Delta$, where $\Delta\approx e^{-\pi}$, and also imply a restriction on the
frequency 
\eqn\freqcond{\omega_\infty r_{max}\le {r_o\over r_{max}}.}
Since $r_0/r_{max}$ is itself constrained to be small by \rzero, equation \freqcond\ tells us how small
$\omega_\infty$ needs to be for our calculation to hold. In the notation of \ms ,
$\winf r_1 \sim O( {r_0 \over r_1})$ corresponds to $ab\sim O(1)$.
Equation \freqcond\  is the condition for $k_5\geq 0$, {\it i.e.}
charge equal to zero, or the same sign as the
black hole, which are the processes with the greatest emission rates. For 
$k_5<0$, the analogue of \freqcond\ is $\mu r_n \leq r_0 /r_{max}$.

\subsec{Matching}
The result for the absorption coefficient in \ms\ comes from matching the
constant and linear terms in the expansions of $\phi_\infty$ \phinfsmal\ and
$\phi_h$ \phhorlarg.
It is important to note, however, that there is
no actual interval in which the asymptotic forms $\phi_\infty$ and $\phi_h$ are
both good approximate solutions.  If we consider the equal charge case,
$r_1=r_5=r_n$, and take $k_5=0$, then $\phi_\infty$
is a good approximation for $r\gg r_1$, while $\phi_h$ is valid for $r\ll r_1$,
giving a clear gap.  Moreover, if we consider the limit in which one of the
charges is much smaller than the other two, such as  $r_n\ll r_1,r_5$ with
$r_1\approx r_5$, then the gap widens further.  The condition for $\phi_\infty$
remains $r\gg r_1$.  However, the condition for $\phi_h$ deduced from \waveinv\
and \decoeff\ becomes $r\ll\sqrt{r_nr_1}$.  Recall that this is the parameter
range in which agreement is expected with the D-brane result. It is clear then,
that, in general, there is a need for a third approximate solution in a middle
region, which can be matched on to $\phi_\infty$ and $\phi_h$ at either end.

Unfortunately, we have been unable to find such an approximate solution and
middle region with overlaps at both ends. However, an alternative way to see the
need for a middle region is to note that the analytic forms for
$\phi_h$ and $\phi_\infty$ in equations \phinfsmal\ and \phhorlarg , which we
want to match, disagree 
functionally\foot{For a discussion of matched asymptotic 
expansions see {\it e.g.} \fluids.}.  
In particular, $\phi_h$
contains a $v\log v$ term for small $v$, while $\phi_\infty$ has a $\log v$
term, and a priori one expects the $\log$'s to be important for small $v$.
In particular, the $v\log v$ term in $\phi_h$ is lower order than the 
$v$ term as $v\rightarrow 0$.
With this in mind, we try a series expansion of the form
\eqn\phmiddle{\phmid = A_0 +A_1 v +... +\log v (B_0 +B_1 v +...),}
which has the appropriate analytic structure to link together the
analytic forms of $\phi_\infty$ and $\phi_h$. 
Substituting $\phmid$ into the wave
equation \waveinv , we find that $\phi_{mid}$ is a good approximate solution for
$r\ll r_{max}$, giving the coefficients $B_n,\ n\ge0$ and $A_n,\ n\ge 2$ in
terms of $A_0$ and $A_1$.  For example,
\eqn\midcoefs{ B_0=C_2A_0+C_3A_1,\qquad B_1= -{C_2^2\over C_3}A_0 -C_2A_1 }
$A_0$ and $A_1$ are then two arbitrary constants determined by the matching.  Matching the constant 
and linear terms between $\phi_h$, $\phi_{mid}$ and $\phi_\infty$ then gives 
\eqn\match{b_2=A_0=AE,\qquad b_1=A_1=AEg.}

Matching the $\log v$ and $v\log v$ terms with those in $\phi_\infty$ and
$\phi_h$ then gives nontrivial consistency checks on the solution, as there
are no free parameters left. One can show that the coefficient of the $\log v$
term in $\phmid$ indeed matches the coefficient $b_3$ in the small argument expansion
of the Bessel function $\phinf$ (see appendix for details). This is good, since
$\log v$ diverges as $v$ approaches zero, and one would, therefore, expect this
term to be important. However, we find that the coefficient $B_1$ of the 
$v\log v$ term in $\phmid$ does not match the corresponding coefficient in the
expansion \phhorlarg\ of the hypergeometric function $\phi _h$. 

Does this lack of agreement pose a real problem?  
The regions of validity for $\phi_h$ and $\phi_{mid}$ overlap for
\eqn\overlap{ \Delta \ll v \ll 1}
If $v$ is in this range, then $|ab v\log v| \ll |gv|$, and it is  reasonable to
keep the linear term and ignore the logarithm in the expansion of the hypergeometric function. 
Still, we should take this as a cautionary note; 
the lack of matching implies that using the form of the hypergeometric
function away from the horizon as given in \phhorlarg\  may not be valid. However, 
the expansion \phhorlarg\ is the key ingredient which yields 
the detailed form of $\abs(\omega)$, which agrees so
well with the D-brane prediction \ms. 

It is our 
view that this matching is correct in the parameter regimes indicated.
As mentioned before, $\phi$ can be expanded in a double power series in $\omega_\infty,\mu$ 
in a middle region. This is a well defined expansion, and
yields a form of $\phmid$ which is the same as in \phmiddle; the zeroth
order term is $\phi \sim a_0 +a_1 v$ and the first order terms
gives the logarithms. The virtue of the expansion is that it is well defined; 
the drawback is that from it one only obtains the very low
frequency form of the absorption coefficient $|{\cal A}(\omega,k_5)|^2\sim \omega \winf ^2 $.
 
Finally, let us complete the calculation of $A(\omega,k_5)$.
The matching conditions \match\ imply that $b_1 /b_2 =g$, which allows us
to solve for $\scat $ in terms of $g$, a known function of the parameters
in the wave equation. This gives
\eqn\scatt{ 1+i\scat =-i {\pi\over 2}\omega ^2 r_0 ^2 g }
Using properties of the digamma function given in the appendix and the relation 
$|\abs |^2 =1- |\scat |^2$, then gives the
expression \clabs\ for the absorption coefficient, 
completing the derivation.

\subsec{Flux Computation}
The conserved particle number current of the Klein-Gordon equation can also
be used to compute the absorption coefficient, as in \refs{\unruh,\ms}.
We use this method to generate a consistency check on our solution,
since the current should be independent of radius $r$. In terms of the tortoise
coordinate $r_*$, the flux is
\eqn\fluxdef{
\flux ={1\over 2i} f^{1/2} r^3 (\phi ^* {\partial \phi \over \partial r_* }
-c.c.) }
Using $\phinf$, \phinfin , this gives as $r\rightarrow \infty $, a sum
of the incident flux and the outgoing scattered flux,
\eqn\fluxinf{ \flux  = \winf (-1 +|\scat |^2 ) }
On the other hand, using the expression \phhorin\ for $\phi$ near the
horizon, the flux is
\eqn\fluxhor{\eqalign{
 \flux =& {1\over 2i} (2 r_0 ^2 (1-v)\phi^* \partial _v \phi-c.c.) \cr
 = & -r_0 ^2 (a+b) |A|^2 \ \  ,v\rightarrow 1 \cr } } %
Now, if one uses the expansion of the hypergeometric function away
from the horizon, equation \phhorlarg , the flux must be the same. Again,
using (A.6) this works out,
\eqn\flhorlarg{\eqalign{
\flux = &-i r_0 ^2 (a+b) |AE|^2 (g-g^* ) ,\qquad v\rightarrow 0 \cr
 = & -r_0 ^2 (a+b) |AE|^2 |E|^{-2} \cr }}
Further, one could use $\phmid$  or the small argument expansion of $\phinf$
to compute the flux. For example, using either \phmiddle\ or \phinfsmal , the
flux in the region $\omega^2r_0^2\ll v\ll 1$ is
\eqn\fluxmid{\eqalign{
 \flux = & -ir_o ^2 (b^*_2 b_1 -b^*_1 b_2 ) \cr
=& -ir_0^2(a+b)|AE|^2(g-g^*).}}
Finally, the absorption coefficient is the ratio of the flux into the
horizon to the incident flux at infinity. From \fluxinf\  the latter
is just $\winf$, and we again get \clabs\  for $\abs$.

\newsec{Agreement and Disagreement with D-Brane Prediction}
We have now established that the spacetime absorption coefficient has the form \clabs,
which we repeat here,
\eqn\absagain{
|{\cal A}(\omega,k_5) |^2 =\omega_\infty ^2 r_0 ^2 \pi ^2 ab
{e^{2\pi (a+b)} -1 \over (e^{2\pi a} -1 ) (e^{2\pi b} -1 ) },}
with constants $a$ and $b$ given by 
\eqn\aandb{\eqalign{ & a=\sqrt{C+D} +\sqrt{D}  ,\qquad  b=\sqrt{C+D} -\sqrt{D}\cr
&C  ={1 \over 4 r_0^2}\left(  \omega_\infty^2 r_1^2r_5^2
 + \mu ^2 (r_1^2r_n^2 +r_5^2r_n^2 ) \right),
\qquad D={\mu^2\over 4r_0^4} r_1^2r_5^2r_n^2, \cr 
& \winf^2= \omega^2 - k_5^2,\qquad \mu = \omega - k_5(1+ r_0^2/r_n ^2)^\half, \cr}}
This result is valid under the conditions 
$r_0^3\ll r_1r_5r_n$ and $\omega_\infty r_{max}\le r_0/r_{max}\ll 1$.
This range of parameters is invariant under U-duality transformations interchanging the three charges.

\subsec{U-duality and Neutral Emission}
For neutral emission ($k_5=0$), the absorption coefficient \absagain\ is itself invariant under
interchange of the three charges $r_1,r_5,r_n$ and in general disagrees with the D-brane result for the 
absorption coefficient in \dbraneabsorp.  We saw in section (2), that we should expect agreement only
over the range $r_n\ll r_1,r_5$.  In \ms\ the spacetime calculation was done restricted to this 
parameter range, yielding remarkable agreement with the D-brane result.  Here we simply show that the 
approximation $r_n/r_1,r_n/r_5\ll 1$ can be made on our final result as well to get agreement with the
D-brane result.

In the neutral case, the coefficients $a,b$ in \absagain,\aandb\ can be put in the form
\eqn\aandbexact{\eqalign{ a=&{\omega r_1r_5r_n\over 2r_0^2}\left(
\sqrt{1+{r_0^2\over r_n^2}+{r_0^2\over r_1^2}+{r_0^2\over r_5^2}}+1\right),\cr
b=&{\omega r_1r_5r_n\over 2r_0^2}\left(
\sqrt{1+{r_0^2\over r_n^2}+{r_0^2\over r_1^2}+{r_0^2\over r_5^2}}-1\right).\cr}}
Making the approximations $r_0,r_n\ll r_1,r_5$ one recovers the result of \ms
\eqn\msresult{\eqalign{
a\simeq & {\omega r_1r_5\over 2r_0^2}\left(\sqrt{r_0^2+r_n^2}+r_n\right)
={\omega\over 4\pi T_R}\cr
b\simeq & {\omega r_1r_5\over 2r_0^2}\left(\sqrt{r_0^2+r_n^2}-r_n\right)
={\omega\over 4\pi T_L},\cr}}
leading to agreement between the exponential factors in \absagain\ and \dbraneabsorp .
The prefactors can also be seen to agree in this limit, 
implying that our expression for the spacetime absorption coefficient agrees overall in this limit 
with the D-brane result, as in \ms.

If instead of $r_n\ll r_1,r_5$, we consider, for example, the U-duality
related limit $r_1\ll r_n,r_5$, then the constants $a$ and $b$ in \absagain\
reduce to
\eqn\msresult{\eqalign{
a\simeq & {\omega r_nr_5\over 2r_0^2}\left(\sqrt{r_0^2+r_1^2}+r_1\right)
={\omega\over 4\pi T_\onebar}\cr
b\simeq & {\omega r_nr_5\over 2r_0^2}\left(\sqrt{r_0^2+r_1^2}-r_1\right)
={\omega\over 4\pi T_1},\cr}}
where the temperatures $T_1,T_\onebar$ were defined above in \onetemp.  It is
then tempting to think of the exponential factors in the denominator in
\absagain\ as arising from Bose-Einstein distributions of 1-branes and
anti-1-branes.  The analogue of the quantized momenta of the left and
right-movers, might then be the number of times a given 1-brane, or
anti-1-brane, is wound.  Perhaps more careful analysis of \absagain\ in this
limit (or the similar limit for 5-branes) could yield insight into
brane/anti-brane dynamics.

\subsec{Charged emission}
Analysis of the charged case is complicated by the difference between the factors 
$\omega_\infty$ and $\mu$.  Validity of the spacetime calculation requires $\omega_\infty r_{max}$ to 
be small compared to $r_0/r_{max}$. However, 
this does not imply that $\mu$ is necessarily small as well.  We consider two cases.
Case I, in which $\omega_\infty^2=(\omega-k_5)(\omega+k_5)$ is small, with both $\omega$ and $k_5$ 
individually small; and Case II, in which $\omega_\infty^2$ is small because either the difference or
sum of $\omega$ and $k_5$ is small, with neither one individually small.  In Case I, 
$\mu=\omega-\sqrt{1+r_0^2/r_n^2}k_5$ is also 
small, of the same order as $\omega_\infty$.  However, in Case II, $\mu$ will be small only if 
$r_0/r_n$ is also small (and it is the difference of $\omega$ and $k_5$, which is small, 
rather than the sum).  

Let us see what effect this has on the coefficients $a$ and $b$, which may be written as
\eqn\chargedcoeffs{\eqalign{a=&{r_1r_5r_n\over 2r_0^2}\left(
\sqrt{\omega_\infty^2{r_0^2\over r_n^2}+\mu^2(1+{r_0^2\over r_1^2}+{r_0^2\over r_5^2})}+\mu\right),\cr
b=&{r_1r_5r_n\over 2r_0^2}\left(
\sqrt{\omega_\infty^2{r_0^2\over r_n^2}+\mu^2(1+{r_0^2\over r_1^2}+{r_0^2\over r_5^2})}-\mu\right).\cr}}

\noindent
{\bf Case I:} $\omega_\infty$ and $\mu$ are the same order of magnitude.  In the limit $r_0,r_n\ll r_1,r_5$,
the $r_0/r_1$ and $r_0/r_5$ terms inside the square root may then be consistently dropped relative to the
other terms and we get
\eqn\toagree{a\simeq{\omega-k_5\over 4\pi T_R},\qquad b\simeq{\omega+k_5\over 4\pi T_L},}
giving agreement between the exponential factors in \absagain\ and \dbraneabsorp, as in \ms.  
Once again the prefactors may also be seen to agree in this limit, giving overall agreement between 
the two expressions. As in the neutral case, the
spacetime and D-brane results disagree for $r_n$ outside of the limit $r_n\ll r_1,r_5$.

\noindent
{\bf Case II:} Assuming that $r_0/r_n$ is not much smaller than 
$1$, then $\mu$ is not small in this case.  The 
$\omega_\infty^2 r_0^2/r_n^2$ term under the square root in $a,b$ then can 
have the same order of magnitude as the
$\mu^2r_0^2/r_1^2,\ \mu^2r_0^2/r_5^2$ terms, 
and the latter may not be consistently dropped relative to the 
former.  The expressions for $a,b$ then do not reduce to 
the form \toagree\ necessary to obtain agreement
with the D-brane result.  With $r_0/r_n\ll 1$, it is simple to see that again
the 
$\omega_\infty^2r_0^2/r_n^2$ term may be of the same order as, or 
less than, the $\mu^2r_0^2/r_1^2,\ \mu^2r_0^2/r_5^2$ terms,
giving disagreement with the D-brane result.

\subsec{Charged Emission Via Boosts}
Another way to analyse the results for charged emission is to exploit, as in \ms,
the boost invariance of the scattering problem.  In \ms\ it was shown that the wave equation
for a scalar, with energy $\omega$ and internal momentum $k_5$, can be rewritten in terms of the
neutral ($k_5=0$) equation, with boosted parameters given by
\eqn\boosted{\omega^{\prime 2}=\omega^2-k_5^2,
\qquad r_n^\prime=r_0\sinh \sigma^\prime , \qquad 
e^{\pm\sigma^\prime}=e^{\pm\sigma }{(\omega\mp k_5)\over \omega^\prime }. }
In terms of the boosted parameters, the coefficients $C$ and $D$ are
\eqn\canddboosted{C={\omega^{\prime 2}\over 4 r_0^2}
\left(r_1^2r_5^2 +r_1^2r_n^{\prime 2} +r_5^2r_n^{\prime 2}\right),
\qquad D={\omega^{\prime 2}\over 4r_0^4} r_1^2r_5^2r_n^{\prime 2} .}
The constants $a$ and $b$ in the absorption coefficient are then 
\eqn\aandbboosted{\eqalign{ a=&  
{\omega^\prime r_1r_5r_n^\prime \over 2r_0^2}\left(\sqrt{{r_0^2\over r_n^{\prime 2}} 
+{r_0^2\over r_5^2} +
{r_0^2\over r_1^2}+1}+1 \right),\cr
b= & {\omega^\prime r_1r_5r_n^\prime \over 2r_0^2}
\left(\sqrt{{r_0^2\over r_n^{\prime 2}} +{r_0^2\over r_5^2} +{r_0^2\over r_1^2}+1}-1 \right)
.\cr }}
In the limit 
\eqn\newapprox{r_0,r_n^\prime\ll r_1,r_5} 
we recover the result \ms ,
\eqn\chargedmsresult{\eqalign{
a\simeq & {\omega^\prime r_1r_5\over 2r_0^2}\left(\sqrt{r_0^2+r_n^{\prime 2}}+
r_n^\prime\right)
={\omega -k_5\over 4\pi T_R}\cr
b\simeq & {\omega^\prime r_1r_5\over 2r_0^2}\left(\sqrt{r_0^2+r_n^{\prime 2}}-
r_n^\prime\right)
={\omega +k_5\over 4\pi T_L}.\cr}}
We then have agreement with the D-brane case, so long as \newapprox\ holds.
We can now ask for what range of the original unboosted parameters \newapprox\ 
fails.  
From the definition of the boosted parameters in \boosted,  we have
\eqn\newsigma{e^{\sigma^\prime}=\sqrt{{\omega -k_5\over \omega +k_5}}e^\sigma,\qquad
e^{-\sigma^\prime}=\sqrt{{\omega +k_5\over \omega -k_5}}e^{-\sigma}.}
From this we see that $r_n^\prime=r_0\sinh \sigma^\prime$ can become large 
(for fixed $r_0$) in two different 
ways, either by $\sigma\rightarrow\pm\infty$, which would make $r_n$ itself large, 
or by having $(\omega -k_5)/(\omega +k_5)$ tend to zero or infinity, keeping $\sigma $ finite.
This latter possibility in which \newapprox\ fails 
is exactly Case II above, in which the sum or difference of $\omega$ and 
$k_5$ is small, with neither individually small.  We have then reproduced the results of the previous
analysis in terms of the boosted parameters.

\subsec{Disagreement in the Fat Black Hole Regime}
In section (2) we argued that the D-brane and spacetime results for the absorption coefficient should agree
over the parameter range $r_0,r_n\ll r_1,r_5$.  We have found, however, that the results for charged 
emission actually disagree within this parameter range, when the quantity 
$\omega_\infty^2=(\omega-k_5)(\omega+k_5)$ is small, without either $\omega$ or $k_5$ being 
small individually.  We now show that this is characteristic of low energy, 
charged emission in the FBH regime.

Recall that in the FBH regime, open string excitations on the brane have $\omega,\ k_5$ 
quantized as integer multiples of $1/N_1N_5R$.  Whereas, modes which propagate in the bulk have 
$k_5$ quantized as an integer multiple of $1/R$.  The lowest energy charged emission possible
would be, {\it e.g.}, a left moving open string with energy and momentum 
$\omega_L=k_{5,L}=(N_1N_5+1)/N_1N_5R$ and a right moving open string with $\omega_R=-k_{5,R}=1/N_1N_5R$,
annihilating to form a closed string with energy and internal momentum 
$\omega=(N_1N_5+2)/N_1N_5R,\ k_5=1/R$.  This process then falls, for $N_1N_5\gg1$, 
within our Case II above, in which the difference between $\omega$ and $k_5$ is much smaller than the
sum (or alternately, sending $r^\prime\rightarrow\infty$), and the spacetime and 
D-brane results disagree.

It should be noted that the size of the disagreement we have found between the spacetime 
and D-brane results is not large\foot{We thank J. Maldacena for emphasizing this point to us.}.  
We have observed above that in the FBH limit, we
can have $r_n^\prime\rightarrow r_1,r_5$, while $r_n$ itself is still small.  
The term $r_0^2/r_n^{\prime 2}$ in \aandbboosted\ is then of the same order as the
$r_0^2/r_1^2$ and $r_0^2/r_5^2$.  The latter terms then cannot be ignored relative to the former.
The exponential terms in the D-brane and spacetime results then differ by terms of this order.
As a check that our spacetime result is accurate to this order, it is straightforward to verify 
that the low frequency limit of \absagain\ yields the black hole area, as the results of \dgm\ demand, 
to this order\foot{An additional order of accuracy in $r_0^2/r_1^2$ can be obtained by considering
the U-duality invariant extension of the results of \km.  See Note Added below.}.

\newsec{Disagreement in Higher Energy Scattering}
The wave equation \wave\ can also be solved for the scattering coefficient $\scat$
at frequencies sufficiently high that the wave is over the scattering
barrier. At these energies, one can use the Born approximation. We will see
that the absorption coefficient rapidly goes to one - this is just
the particle limit, in which the particle is captured. However, the D-brane
prediction for the emission rate, still given by \rate\ and no
longer matches the Hawking emission in this higher frequency part of the
spectrum. For classical sized black holes, these higher frequencies
are still well below the string scale, at which corrections to the 
approximate vertex \refs{\hashimoto,\dmtwo} used in the computation of \dbraneabsorp\ should 
be important.

First rewrite the wave equation in a standard scattering form. The
tortoise coordinate is defined by
\eqn\tortoise{d r_* ={\sqrt{f}\over 1-{r_0 ^2 \over r^2} }dr }
Let  $ \lambda =r^{3/2} f^{1/4}$ and let $\chi = \lambda \phi$. Then
the wave equation \wave\  becomes
\eqn\wavetort{\chi ''(r_* ) +\left[ \winf ^2 -V_{coul} -V_{grav} \right]
\chi =0, }
where
\eqn\scatpot{V_{coul} =r_n ^2 {\winf ^2 -\mu ^2 \over r^2 +r_n ^2 },\qquad  
V_{grav} ={\lambda '' (r_* )\over \lambda }, }
and prime denotes differentiation with respect to $r_*$.
The total potential falls off like $r_* ^{-2}$ as 
$r_* \rightarrow\infty$, so the parameter $\winf$ is the frequency at infinity.
The ``Coulombic'' potential is monotonic decreasing from the horizon
to infinity. The ``gravitational'' potential falls off exponentially
fast towards the horizon ($r_*\rightarrow -\infty$) like $ e^{2\kappa r_*}$, where $\kappa =
2\pi T_H$ is the surface gravity.

Near the horizon \wavetort\  becomes
\eqn\wavtorhor{\chi '' +\omega _h ^2 \chi \simeq 0 ,}
up to exponentially decaying terms, where $\omega _h =\omega 
-k  (1+{r_0 ^2\over r_n ^2 })^{-1/2}$. This verifies the claim
made earlier that near the horizon the solutions are plane waves
in the tortoise coordinate, with frequency $\omega _h$.

The absorption coefficient \clabs\ describes the scattering behavior
of the solutions to \wavetort\  in the low frequency limit $\winf r_{max} \ll 1$.
At high frequencies when most of the incident wave is absorbed, the Born
approximation can be used: $| \scat |^2 =  {1\over 4\winf ^2} | \int dy V(y)
e^{-2i\winf y} |^2$, valid when $\winf^2 > Max (V)$. To
find the frequency cutoff, we need the height of the potential
barrier. We specialize to the neutral case at this point.  In this case $V_{coul}$ vanishes.
The result in the charged case is similar.

The gravitational potential $V_{grav}$ is rather complicated and is 
difficult to 
maximize analytically. However, it is straightforward to estimate
the height. One finds that $Max (V)\sim O(1) r_{max} ^{-2}$, and
so we need
\eqn\overone{\omega>\omega_{over}\simeq 1/r_{max}}
Once this condition on the frequency is satisfied, the integral in
the Born approximation above goes rapidly to zero with increasing
$\omega$: the integrand is an oscillatory factor times the positive 
definite potential, and one can check that the width of the potential
is much greater than the wavelength. Therefore
\eqn\highabs{ |\abs(\omega)|^2 \rightarrow 1,\qquad \omega>\omega_{over}}
This classical result \highabs\ clearly disagrees with the D-brane 
prediction \dbraneabsorp.

It is simple to check that $\omega_{over}$ is above the Hawking temperature, their relation 
being given approximately by
\eqn\overit{{\omega_{over}\over T_H}\simeq {2\pi r_{max}r_n\over r_0^2}\gg 1.}
Emission is therefore going to zero in this regime.  It seems reasonable, however, to expect emission 
from the D-brane system to approach zero in the same manner.  We note that even though 
$\omega_{over}\ll T_H$, it may still be small compared to the mass difference $\Delta M=N_R/R$
from extremality.  We find that $\omega_{over}\ll \Delta M$ is satisfied if
\eqn\smallenough{r_{max}\gg {16 g^2 r_n^2\over RV r_0^4}.}
In this case it is appropriate to model emission as a thermal process.
The absorption coefficient therefore approaches unity below the energy range in which the considerations
of reference \refs{\esko,\dasgupta} are necessary.

There appears to be an error then in the D-brane prediction
at high frequencies (but frequencies which are still small compared
to the string scale, for macroscopic sized black holes). The basic
vertex used in the D-brane calculations 
does not correctly describe the higher energy processes.
This vertex is the first term in a low-energy expansion of 
the exact string vertex in \hashimoto . However, it is not difficult to 
check that, if one considers the
next terms in the expansion, the behavior of the resulting 
approximate vertex also does not give an absorption coefficient
approaching unity. If the D-branes are to give this characteristic
black hole behavior in the particle limit, there is some piece
of D-brane physics which is missing from the present picture. 

\newsec{Conclusions}
In this paper, we have shown that the spacetime absorption coefficient
has the form found in \ms\ and given in \clabs\ over a large range of parameter space.
Moreover, our derivation gives explicit limits to the range of validity of this result.
The spacetime result \clabs\ disagrees in general with the D-brane expression \dbraneabsorp.
We argued in section 3, based on the conjectured statistical mechanics of 
D-branes, that agreement should be expected for $r_n\ll r_1,r_5$.  

In section 5, we saw that in the neutral case, the results agree over precisely this regime 
(as we would expect, based on \ms).  However, for charged emission, we found that the expressions
disagreed for parameters in the FBH regime, in which charge is quantized in much smaller 
units on the brane than in the bulk of spacetime.  Whether, or not, 
one accepts the argument of section 3, this demonstrates that the current D-brane calculations are at 
best insufficient to describe charged emission in the FBH regime correctly.  It seems plausible
that a better understanding of D-brane dynamics will resolve this issue, 
insofar as
in the current model the difference in charge quantization conditions, between the 
brane and the bulk of spacetime, prohibits low lying left-moving excitations on the brane from 
contributing to charge emission.

In section 6, we showed that the Born approximation can be used to show that for 
$\omega>\omega_{over}$, but still below the string scale, 
the spacetime absorption coefficient approaches unity, as it should in the particle limit.  
This behavior, however, is not seen in the D-brane results.  The effective vertex used in the
D-brane calculation should be valid up to the string scale.  This result then poses an additional
challenge for the D-brane model.

\bigskip\noindent
{\bf Note Added:}
After this work was completed the paper \taylor\ appeared also giving the
expression for the spacetime absorption coefficient \clabs\ over the extended 
parameter region.  

In addition, the paper \km\ appeared which treats the case
of two charges, {\it e.g.} $r_1,r_n$, being small.  
In attempting to reconcile our result 
with that of \km\ we discovered that, in the limit 
\eqn\kmlimit{r_1,r_n,r_0\ll r_5,} 
the requirement 
for the hypergeometric equation to hold becomes $r\ll r_0$, implying that our result 
does not hold in this regime.  Fortunately, the methods of \km\ lead simply to an 
improved U-duality invariant result.  Klebanov and Mathur showed that the $C_2$ term in 
\waveinv\ may be included in the mapping to the hypergeometric equation.  Carrying
through the calculation in \km\ without making the approximation \kmlimit\ then gives a
result for the absorption coefficient of the form \clabs\ but with $\sqrt{C+D}$ 
replaced by $\sqrt{C+D+C_2}$ in the coefficients $a$ and $b$ in \clabs.  It is interesting to note
that if the result were extended to $\sqrt{C+D+C_2+C_3}$, then the exponential factor in
the numerator of \clabs\ would exactly cancel the Hawking term.  Moreover, this modification also
correctly yields the exact horizon area in the low energy limit of the absorption coefficient and fits in
well with the considerations of \larsen .

\bigskip\noindent
{\bf Acknowledgements:}
We would like to thank J. Gauntlett for useful conversations.
DK and JT thank the Aspen Center for Physics, where this project was initiated.
FD thanks the Relativity group at DAMTP, Cambridge, for hospitality during the
writing up of this work.
JT is supported in part by NSF grant  NSF-THY-8714-684-A01.
FD was supported in part by US DOE  grant DE-FG03-92-ER40701 
at the Lauritsen Laboratory, Caltech. 

\appendix{A}{Details of Absorption Coefficient Calculations}

The solution to the wave equation \waveinv\  is constructed by matching
three pieces. Near infinity ($v=0$), the solution is given by a Bessel function
\phinfin , near the horizon ($v=1$) by a hypergeometric function \phhor , and
these are linked together by a ``middle solution'' \phmiddle . The large
$v$ (small $r$) expansion of the Bessel function can be matched
to $\phmid$, as is done in \match . The large $v$ expansion of the
Bessel function is given in \phinfsmal , where the coefficients are

\eqn\besscoef{b_1 =i \sqrt{ {2 \over \pi }}e^{-i\pi /4} {(1+ i\scat )
\over r_0 ^2 \sqrt{\winf }} 
, \qquad  b_3 = i { e^{-i\pi /4}\over 2\sqrt{2 \pi }}\winf ^{3/2}
(1+i\scat ) ,}
\eqn\btwo{
b_2 ={1\over 2} e^{-i\pi /4}\sqrt{ {\pi \over 2}}\winf ^{3/2} 
\left [ (1-i\scat ) + {i\over \pi} (1+ i\scat )( 1-2\gamma +2 \ln 2 -\ln \winf ^2
r_0 ^2 )\right ] .} 

Likewise, the  hypergeometric function can be expanded for $v\rightarrow
0$. To derive \phhorlarg  , one can use, as in \ms , equation (15.3.6) from 
Abramowitz and Stegun\sa . (There is a misprint in (4.21) of \ms; the first two
arguments of the last hypergeometric function should be interchanged.)
This involves inserting a regulator and taking a limit. Alternatively,
one can directly use equation (15.3.11) from \sa , which gives
\eqn\expf{F(-ia ,-ib, 1-ia -ib ,1-v )\rightarrow E\left[ 1+ v (g - 
{i\over 2}(a+b) ) -ab v\ln v \right],\qquad  v\rightarrow 0 }
where $g$  is given in \ggg  , and as found in \ms , 
\eqn\eee{E(a,b) \equiv {\Gamma (1-ia -ib )\over \Gamma (1-ia) \Gamma
(1-ib)} .}
In deriving the form \clabs for $|\abs |$ one uses (6.1.31) of
\sa\  and finds, as in \ms ,
\eqn\esquared{ {1\over |E|^2} ={2\pi ab \over (a+b)} {(e^{2\pi (a+b)} -1 )
\over (e^{2\pi a} -1 )(e^{2\pi b} -1 )} }
In working out some of the bounds, and when computing the flux, we use
some of the properties of $g$, defined in \ggg . Using (6.3.13) of \sa\
one has
\eqn\img{\eqalign{
 Im (g) = & { a+b \over 2} -ab Im \left[ \psi ( 1-ia) +\psi (1-ib) \right]\cr
=&\pi ab   {(e^{2\pi (a+b)} -1 ) \over (e^{2\pi a} -1 )(e^{2\pi b} -1 )} \cr }}
And, using (6.3.17) of \sa  ,
\eqn\reg{Re \left(g (a,b)\right)=ab\left[ 1 -a^2 \sum_{n=1} {1\over n}
{1\over n^2 + a^2 }  -b^2 \sum_{n=1} {1\over n} {1\over n^2 + b^2 }
\right] }
So over the range of interest for  $a,b$, we have
\eqn\regone{ Re \left(g(1,1)\right) \simeq -{1\over 5} }
\eqn\regsmall{Re \left(g(a,b)\right) \simeq ab,\qquad  a,b \ll 1 }

Lastly, we make some comments about the solution in the middle region
and the matching. When the ansatz \phmiddle\  for $\phmid$ is
substituted into the wave equation \waveinv , this form of 
solution requires that
terms of order ${C_3 a_0 \over v^3}$ have been dropped compared
to ${C_2 a_0 \over v^2}$. This is consistent for 
${r_0 ^2 \over r_1 ^2}\ll v $. Although this requirement looks like we are
losing the overlap region, this does match onto the Bessel function.
This is most easily seen by
looking at how the analogous expansion works for Bessel's equation:
Bessel's equation for the
function  $f=r^{-3/2} H^{(1,2)}$ written in the variable $v$
is $f''(v) +( {C_3 \over v^3 } +{C_2 \over v^2 })f =0$. Substituting in
an expansion of the form \phmiddle\  gives the same relation for 
$B_0$ as in \midcoefs.

On the other hand, it can also be seen directly from the differential equations
that the coefficient $B_1$ of the $v\ln v$ term for the actual solution, 
is different from the coefficient inferred from the hypergeometric equation.

\listrefs
\end